\newcommand{\invnb}{nb$^{-1}$}
\documentclass{PoS}

\title{Operation of the ATLAS detector with first collisions at 7 TeV at the LHC}

\ShortTitle{Operation of the ATLAS detector with first 7 TeV collisions}

\author{\speaker{Peter Onyisi}, on behalf of the ATLAS Collaboration\\
        University of Chicago\\
        E-mail: \email{ponyisi@hep.uchicago.edu}}

\abstract{The ATLAS experiment has successfully recorded over 300 \invnb\ of $pp$ collisions at 7 TeV provided by the Large Hadron Collider, with an efficiency of 94\%.  We describe the data acquisition, trigger, reconstruction, calibration, monitoring, and luminosity measurement infrastructure that have made this possible.
}

\FullConference{35th International Conference of High Energy Physics - ICHEP2010,\\
		July 22-28, 2010\\
		Paris France}

\begin{document}

\begin{figure}
\includegraphics[width=.5\linewidth]{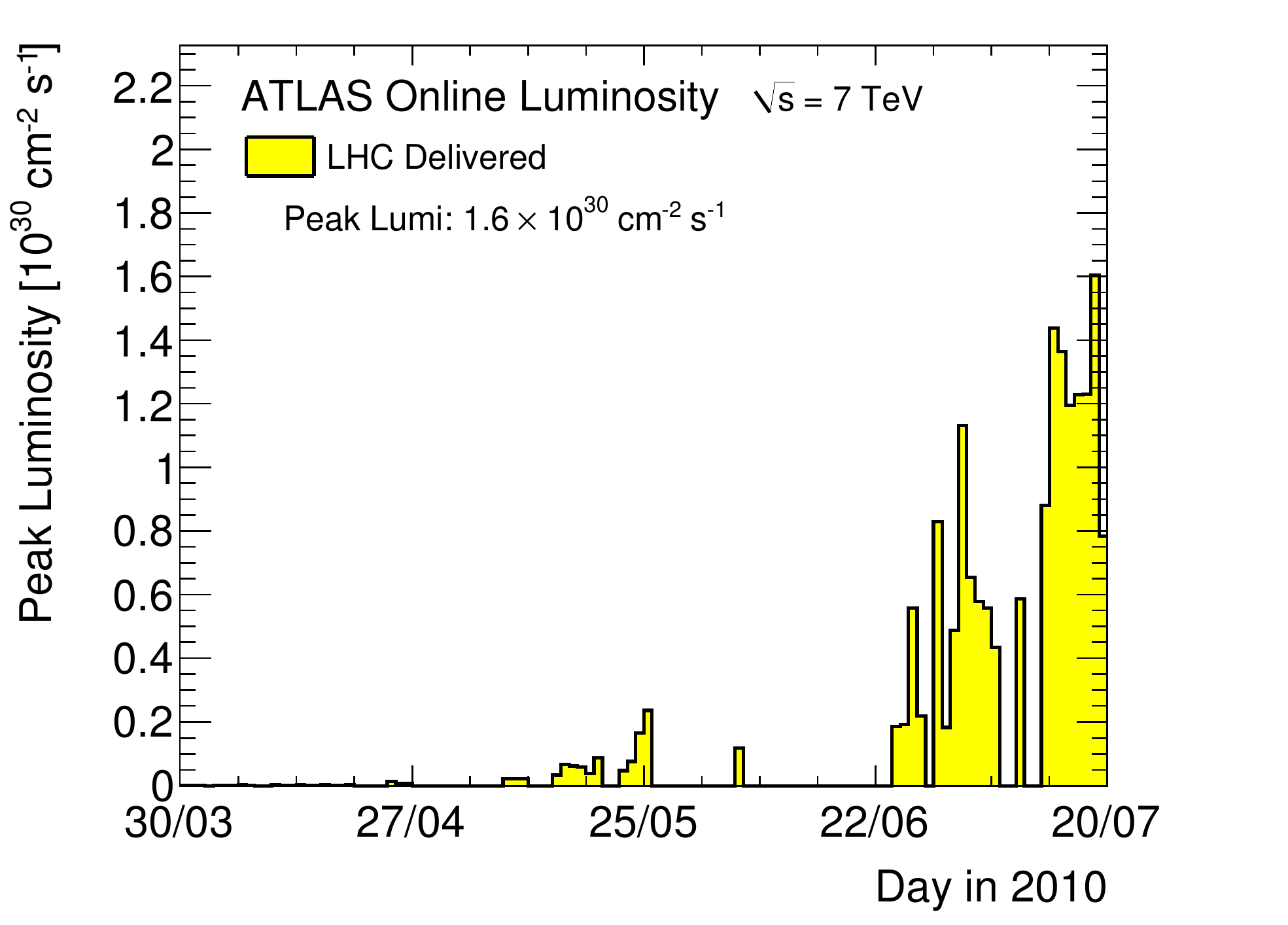}%
\includegraphics[width=.5\linewidth]{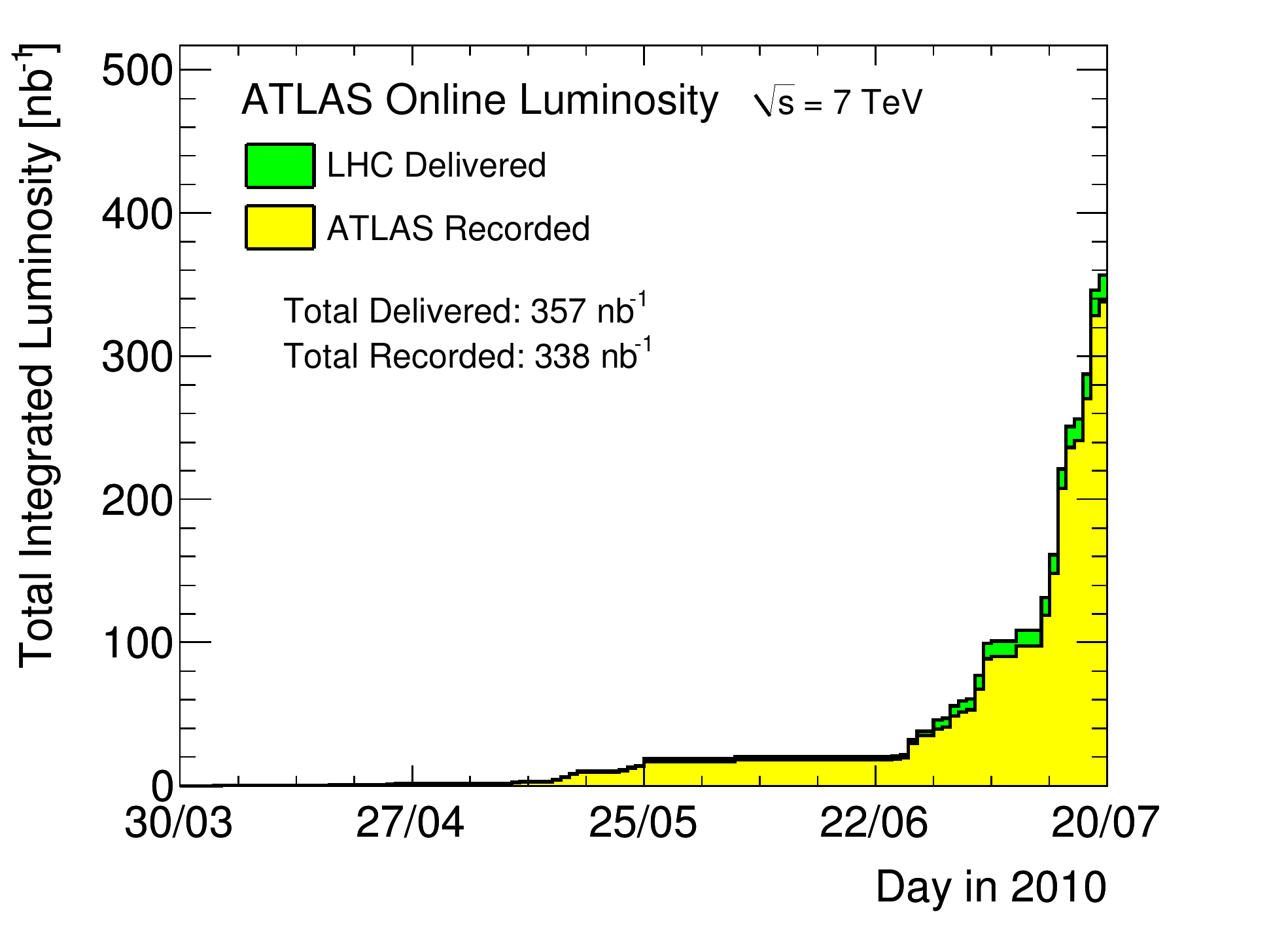}
\caption{\label{fig:profile} LHC luminosity profile as seen by ATLAS through July 19, 2010.  Left: peak luminosity; right: integrated luminosity delivered to ATLAS (green) and recorded by ATLAS in physics configuration (yellow).}
\end{figure}

\section{Introduction}
The ATLAS experiment at the Large Hadron Collider (LHC) is being commissioned with the first $pp$ collisions delivered at 7 TeV.  In these proceedings we describe luminosity measurement, data acquisition and trigger, monitoring, and offline calibration and reconstruction of recorded events to prepare them for physics use.

\section{Luminosity Measurement}
\begin{figure}
\includegraphics[width=.5\linewidth]{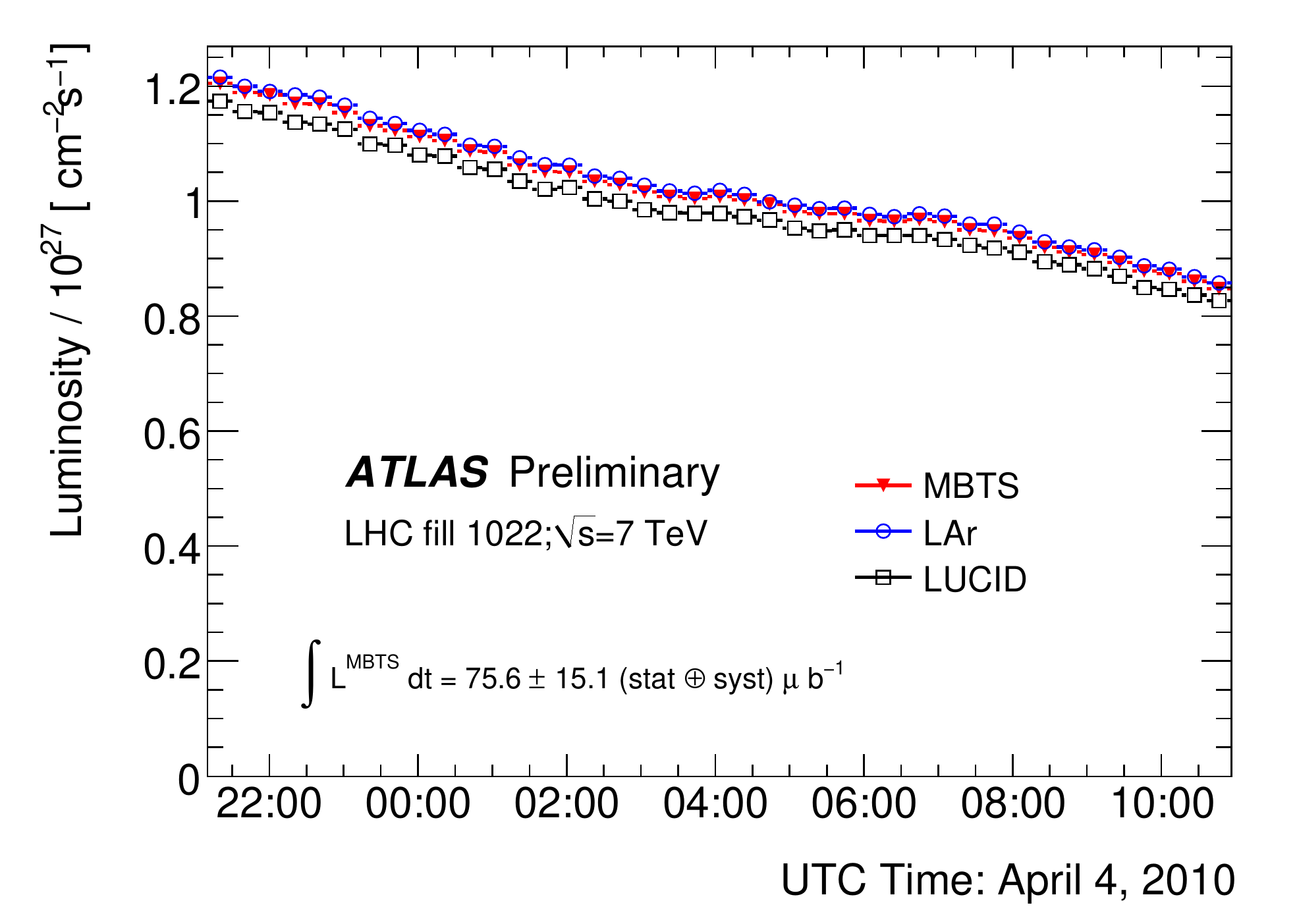}%
\includegraphics[width=.5\linewidth]{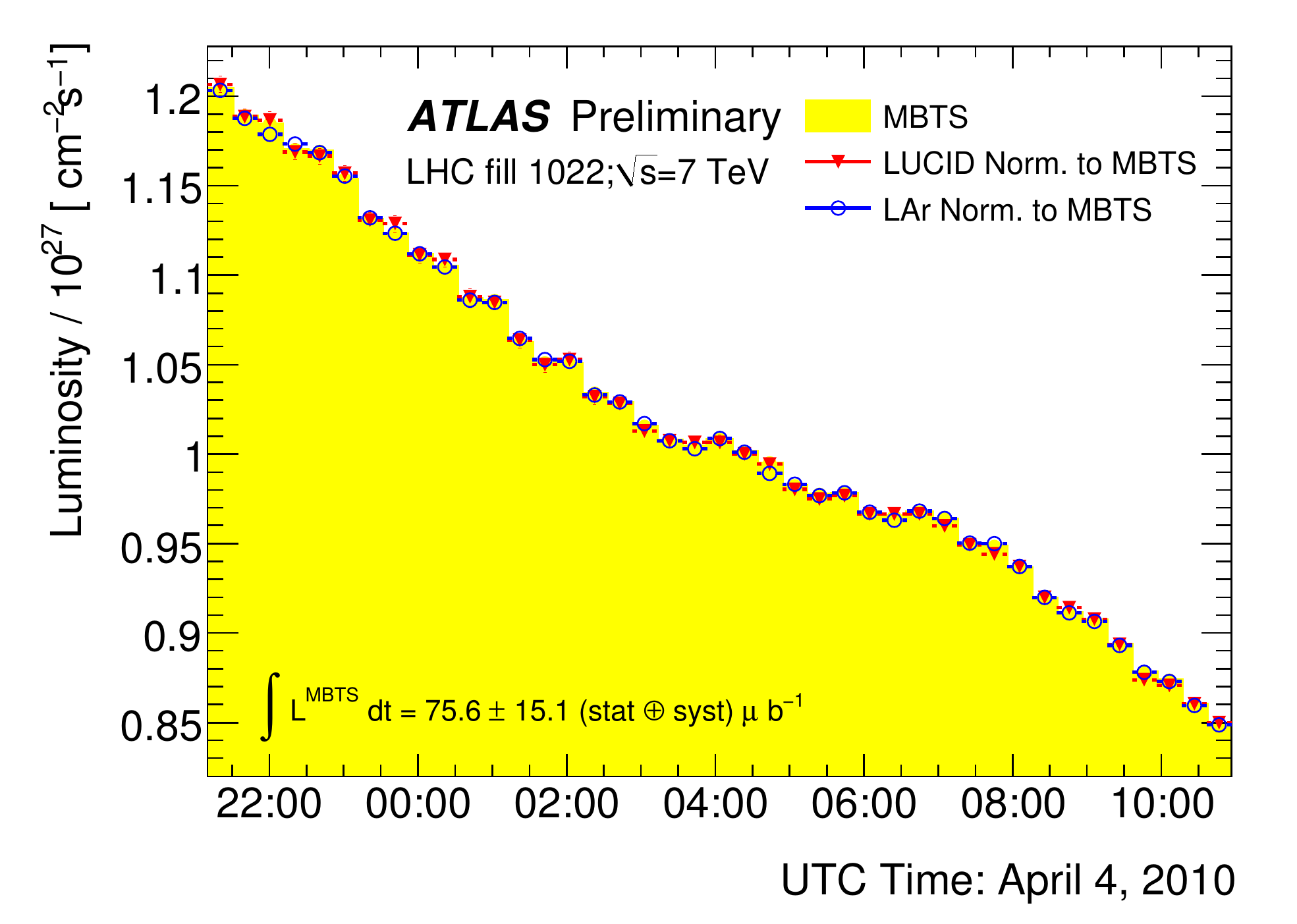}
\caption{\label{fig:lumi} Relative performance of three ATLAS luminosity measurements.  Left: all three measurements are independently normalized to Monte Carlo simulations; differences between the points reflect acceptance and efficiency differences between simulation and data.  All points have a correlated systematic uncertainty (not shown) due to the minimum bias cross-section.  Right: the three measurements for the same fill, normalized to the MBTS luminosity.  The three measurements track each other closely.}
\end{figure}

Luminosity measurement (Fig.\ \ref{fig:profile}) at ATLAS can be done using several subdetectors.  The primary sources of information used so far have been the minimum bias trigger scintillator disks (MBTS), a double-arm Cherenkov detector (LUCID), and the forward liquid argon calorimeter (LAr).  Measurements from these three systems agree very well with each other (Fig.\ \ref{fig:lumi}).

The smallest unit of time for which luminosity is determined in ATLAS is a {\it luminosity block}, during which detector conditions are not supposed to change.  The default length of a luminosity block is adjustable and was set to two minutes for 7 TeV $pp$ running.  A number of conditions, such as a change in trigger configuration, will cause the start of a new luminosity block.

To translate a rate measurement from a luminosity detector into an absolute luminosity, a detector normalization is required.  Initially these were taken from Monte Carlo simulations, and the systematic uncertainty on the assumed interaction cross-section dominated the total 20\% luminosity uncertainty.  To improve this situation, the LHC provided three beam separation scans \cite{vdM}, from which the absolute luminosity could be determined from the geometrical beam profiles and the beam current intensity.  These scans have reduced the total luminosity uncertainty to 11\%, dominated by 10\% due to the beam intensity measurement (5\% per beam, completely correlated between them).  More details may be found in ref.~\cite{lumi}.

\section{Data Acquisition and Trigger}

The ATLAS Trigger and Data Acquisition systems \cite{daq} are designed to perform event selection and readout of the peak 40 MHz bunch crossing rate eventually to be provided by the LHC, reducing it to an output rate of $\approx$ 200 Hz of events most relevant for physics analysis.  

The DAQ system pipelines the detector readouts while awaiting trigger decisions, builds events from the selected interactions, and passes them on to the output.  The final output bandwidth of 200 Hz is constrained by offline reconstruction speed and transfer to mass storage and is not a hard limit in the DAQ.

The trigger comprises three levels: a hardware Level 1 and software Levels 2 and 3 (the third level trigger is referred to as the {\it Event Filter} (EF), and Level 2 and the EF are collectively called the {\it High Level Trigger} (HLT)).  The Level 1 trigger takes input from the calorimeters and muon detectors (as well as some other systems such as the MBTS).  Precisely timed trigger signals allow Level 1 to determine which bunch crossing the event occurred in, and the LHC beam configuration is used to determine the category (colliding bunches, unpaired bunch, empty, \ldots) of the bunch crossing.  Level 1 also establishes Regions of Interest (RoIs) which allow the Level 2 software to only consider a limited fraction of the detector information around the trigger objects.  The EF is able to consider the full event when coming to a decision.  The HLT software has access to information from all subdetectors, including in particular the inner tracking detectors.  Based on the passed triggers, an event will be sent to one or more {\it streams} (muon, jet, electron/photon, \ldots). A typical run may have output rates of Level 1, 2, and EF of 2--3 kHz, 0.7--1 kHz, and 200 Hz respectively.  Currently only a third of the projected HLT farm capacity is installed; more nodes will be added as necessary.  More information on the commissioning of the trigger algorithms is available in other contributions to these proceedings.

The DAQ implements several techniques to improve data-taking efficiency, such as the {\it stopless removal} and {\it recovery} procedures, in which selected detector readouts can be removed if they are causing deadtime and later re-included once fixed.  An automated {\it expert system} is able to recognize specific conditions and suggest actions to the operator or carry them out itself.  

The expert system is used in the {\it warm start} and {\it stop} procedures, in which sensitive detector elements (the silicon tracking detectors and the muon detectors) are held in a standby state until the LHC operators declare the beam to be stable.  During this Standby period a modified trigger menu, focused on monitoring, is active.  The detectors are then brought to their operating state and the trigger menu is reconfigured to physics mode, after which ATLAS is declared Ready and data taking resumes.  The reverse occurs for the warm stop.  These allow ATLAS running to proceed safely while reducing the latency at the beginning and end of stable beam periods.

The DAQ system has functioned very well for initial data taking, with a luminosity-weighted live time of 97\%.  Inefficiencies come from the periods during which ATLAS transitions from Standby to Ready and vice versa, as well as detector readout problems and the stopless removal/recovery procedure.

\section{Online Detector Monitoring}
A comprehensive suite of tools is available, both online during data taking and afterwards, for monitoring detector performance and the quality of the data.  Sources of online monitoring information include the Detector Control System and LHC interfaces, trigger rate monitoring and internal monitoring in the HLT \cite{trigmon}, inspection of raw data fragments, and full event reconstruction using offline code adapted for the online environment.  These sources are available to the personnel in the control room via local applications \cite{dqmd} and are also accessible via web-based interfaces to remote users.  Automated tools compare histograms to expectations and note anomalies.  Histograms are archived during and after a run.

\section{Prompt Reconstruction and Calibration}
\begin{figure}
\begin{center}
 \includegraphics[width=.7\linewidth]{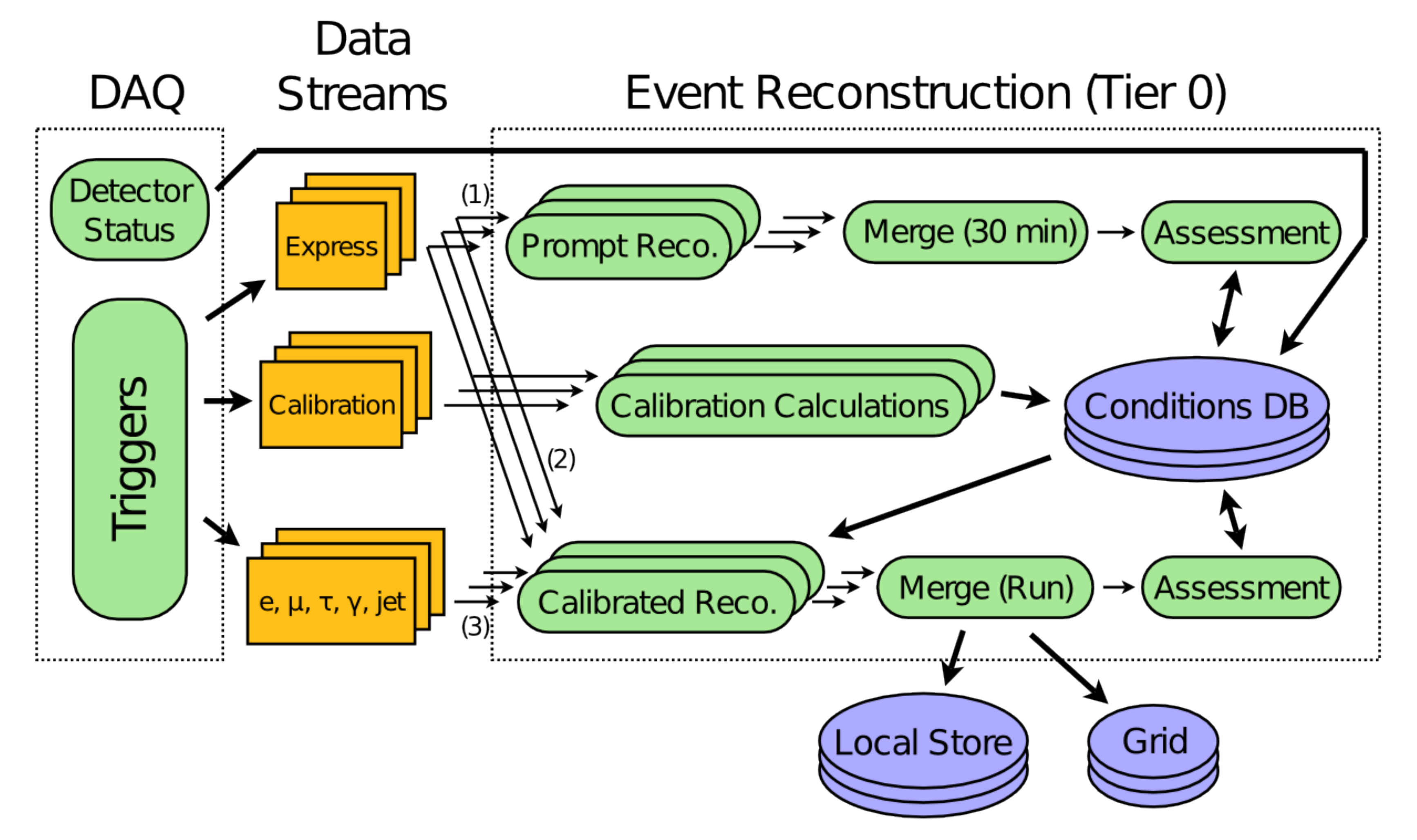}
\caption{\label{fig:dataflow}Schematic of the offline reconstruction of events.}
\end{center}
\end{figure}

The {\it Tier-0} reconstruction of ATLAS data (Fig.\ \ref{fig:dataflow}) that occurs at CERN involves three main phases: a first ({\it prompt}) reconstruction pass on a subset of the data, computation of calibrations from this pass, and a second ({\it bulk}) reconstruction pass.  The goal of this procedure is to provide analysis-quality data to the collaboration within 3--4 days of acquisition.

As files are copied to mass storage at CERN from the DAQ, the Tier-0 batch farm at CERN schedules prompt reconstruction jobs on the raw data.  The prompt reconstruction runs on the {\it express} data stream, consisting of around 10 Hz of events used for monitoring, as well as certain additional calibration streams which are useful for studying noise.  The prompt reconstruction provides the standard analysis input files, as well as monitoring histograms which are made available via web pages.  This prompt reconstruction can be used for calibration purposes such as flagging of noisy cells in the calorimeters.  In addition more minimal reconstruction jobs dedicated to specific systems are run in parallel, which determine for example the beamspot position and width at 10 minute intervals.  The calibrations are then uploaded to the conditions database.  Thirty-six hours after the end of a run, the bulk reconstruction starts with the updated calibrations, reconstructing all data streams (including those processed in the prompt reconstruction, for comparison purposes).  The conditions database ensures that the prompt and bulk reconstruction calibrations are kept separate, and prevents the modification of calibrations for runs already taken or being reconstructed.  The results of the bulk reconstruction are then distributed to Grid sites.

During early data taking Tier-0 reconstruction output has been used for analysis; this requires that as much as possible the software be compatible for all run periods.  This has been enforced with the ``frozen Tier-0 policy'' which only allows changes that have no effect on physics output (such as fixes for crashes in specific events or improvements in monitoring code).  A test suite is used to verify compliance.


\section{Data Quality Assessment}
In parallel with the prompt reconstruction and calibration loop, an assessment is made of the quality of the data just taken.  Initial input to this assessment comes from automated checks on the DAQ configuration, the detector control system status \cite{dcsd}, histograms monitored both online and offline, and the online shifters for each system.  Preliminary decisions are made within 36 hours of the end of a run and are generally finalized within 96 hours of data collection.

A typical run in ATLAS is heterogeneous: it begins before beam injection and ends after beam dump; it features various trigger configurations; ATLAS will be moved from Standby to Ready status and back again; and due to stopless removal and recovery, the portions of the detector that are read out may change.  There are also detector problems, such as power supply trips, that are localized in time.  To handle these effects, the DQ assessment is made with luminosity block granularity.

Typical problems encountered in the DQ assessment include power supply trips, noise bursts, and stopless removal of parts of the detector.  Each subdetector recorded good data for $\gtrsim 94\%$ of delivered LHC stable beam luminosity (Table~\ref{tbl:dqeff}).

\begin{table}
\begin{center}
 \begin{tabular}{ccccccccccc}
\hline\hline
  \multicolumn{3}{c}{Inner Tracking} &
  \multicolumn{4}{c}{Calorimeters} &
  \multicolumn{4}{c}{Muon Detectors}\\
\multicolumn{3}{c}{Detectors}\\
\hline
Pixel & SCT & TRT & LAr  & LAr  & LAr & Tile & MDT & RPC & TGC & CSC\\
      &     &     & EM   & HAD  & FWD     \\
\hline
97.1  &98.2 & 100 & 93.8   & 98.8    & 99.1    & 100  & 97.9 & 96.1 & 98.1 & 97.4\\
\hline\hline
 \end{tabular}
\caption{\label{tbl:dqeff} Luminosity weighted relative detector uptime and good quality data delivery during 2010 stable beams at $\sqrt{s}=$ 7 TeV between March 30th and July 16th (in \%).}
\end{center}
\end{table}

\section{Conclusion}
The ATLAS detector has begun taking data with 7 TeV collisions successfully and with high efficiency.  Commissioning of all essential data-taking systems is proceeding well and calibration and review procedures are capable of producing analysis-quality data within $\sim$ 4 days of data taking.

\end{document}